\documentclass[twocolumn,epjc3]{svjour3}  
\smartqed  
\RequirePackage{graphicx}
\journalname{Eur. Phys. J. C}
\usepackage{amsmath,amssymb,calc}
\usepackage[bulletsep]{collref}
\usepackage{bbm}
\usepackage{bm}
\usepackage{statmath}
\usepackage{braket}
\usepackage{slashed}
\usepackage{bbold}
\usepackage{xcolor}
\newcommand{\matel}[3]{\langle #1|#2|#3\rangle}
\usepackage{hyperref}

\usepackage[final]{feynmf}

\def\beq{\begin{equation}}
\def\eeq{\end{equation}}

\newcommand{\beqn}{\begin{eqnarray}}   
\newcommand{\eeqn}{\end{eqnarray}}

\newcommand{\gsim}{\lower.7ex\hbox{$
\;\stackrel{\textstyle>}{\sim}\;$}}
\newcommand{\lsim}{\lower.7ex\hbox{$
\;\stackrel{\textstyle<}{\sim}\;$}}

\numberwithin{equation}{section}

\begin{document}
\title{OPE-based Methods in Nonperturbative QCD}
\subtitle{SVZ sum rules, \boldmath{$1/M_Q$} expansion, and all that}

\titlerunning{OPE-based Methods in Nonperturbative QCD}        

\author{Mikhail Shifman}

\authorrunning{Mikhail Shifman}

\institute{
William I. Fine Theoretical Physics Institute, University of Minnesota, Minneapolis, MN 55455, USA\\
\email{shifman@umn.edu}
}

\date{FTPI-MINN-22-23, UNM-TH-4132/22}

\maketitle

\begin{abstract}
  I describe the inception and development of nonperturbative OPE-based methods in hadronic physics, such as the SVZ sum rules, inverse heavy quark mass expansion (IHQME) and so on and related topics. 
  
  {\em Invited contribution to the EPJC Volume celebrating 50 years of Quantum Chromodynamics, to be published in December 2022. }
  
\end{abstract}


%

\section{Preamble}

Rewind to autumn of  1971. I am a student at ITEP in Moscow, working on my Masters degree.  The  famous paper of Gerhard 't Hooft \cite{GtH1}
was published in Nuclear Physics in October, but neither myself nor anybody else in ITEP immediately noticed this ground-breaking publication. 
At that time I did not even know what Yang-Mills theories meant. Now, when I think of inception of QCD, the memories  of this paper and its sequel \cite{GtH2}
(issued in December of 1971) always come to my mind. For me, psychologically this was the beginning of the QCD era. 

To give an idea of the scientific atmosphere at that time I looked through the Proceedings of the 1972 International
 Conference On High-Energy Physics \cite{HEP1972}. Theoretical talks were devoted to dual models (a precursor to string theory), deep inelastic scattering and Bjorken scaling, current algebra,
 $e^+e^-\to$ hadrons, etc. In  three talks -- by Zumino, Bjorken and Ben Lee -- the Weinberg-Salam model (a precursor to the present-day Standard Model)  was reviewed.\footnote{There is a curious anecdote I heard later: In December 1979, after the Glashow-Weinberg-Salam Nobel Prize ceremony, a program was aired on Swedish radio. At some point, Weinberg quoted a phrase from the Bible. Salam remarked that it exists in the Quran too, to which Weinberg reacted: ``Yes, but we published it earlier!"}
Ben Lee was the only person to refer to 't Hooft's publications \cite{GtH1,GtH2}. The last talk of the conference summarizing its major topics  was delivered by Murray Gell-Mann. In this talk Gell-Mann discusses, in particular, whether quarks are physical objects or abstract mathematical constructs. Most interesting for us is his analysis of the $\pi^0\to 2\gamma$ decay.
Gell-Mann notes that if quarks are fermions then theoretically predicted amplitude is factor of 3 lower than the corresponding experimental result, but makes no statement of the inevitability of the quark color.\footnote{For me personally the following remark in his talk was a good lesson for the rest of my career:  ``Last year the rate of $K_L^0 \to \mu^+\mu^- $ decay was reported  to be lower than allowed by unitarity unless fantastic hypotheses are concocted. Now the matter has become experimentally controversial." Alas... concocting fantastic hypotheses was the core of my Masters thesis.}

In October 1972 I was accepted to the ITEP graduate school. My first paper on deep inelastic scattering in the Weinberg-Salam model was completed in early 1973; simultaneously, I started studying Yang-Mills theories (in particular, the Faddeev-Popov quantization \cite{FP})  in earnest. At the same time, somewhere far away, behind the Iron Curtain, Callan and Gross searched for a theory with an ultraviolet fixed point at zero. In July of 1973 Coleman and Gross submitted to PRL a paper  asserting that ``no renormalizable field theory that consisted of theories with arbitrary Yukawa, scalar or Abelian gauge interactions could be asymptotically free"  \cite{DG1}.
Damn Iron Curtain! If Gross asked anyone from the ITEP Theory Department he would have obtained the answer right away. The above theorem was 
known to the ITEP theorists from the Landau time. For brevity I will refer to it as the Landau theorem, although it was established by his students rather than Landau himself.
The general reason why this theorem holds was also known -- the K\"allen-Lehman (KL) representation of the polarization operator plus unitarity. 

An explanatory remark concerning the Landau theorem might be helpful here. For asymptotic freedom to take place the first coefficient of the $\beta$ function must be {\em negative}. The sign of the one-loop graphs which determine the coupling constant renormalization is in one-to-one correspondence with the sign of their imaginary parts (this is due to the dispersion KL representation for these graphs). Unitarity implies the positivity of the imaginary parts which inevitably leads to the {\em positive}  first coefficients in the 
$\beta$ functions in renormalizable four-dimensional  field theories based on arbitrary Yukawa, scalar or Abelian gauge interactions. This situation is that of the Landau zero charge in the infrared rather than asymptotic freedom. In Yang-Mills theories in physical ghost-free gauges some graphs have no imaginary parts which paves the way to asymptotic freedom (see e.g. \cite{MS1}).

In fact, it is quite incomprehensible why asymptotic freedom had not been discovered at ITEP after 't Hooft's 1971 publication. In Ref. \cite{MS1} the reader can find a 
 narrative  about this historical curiosity.

May 1973 should be viewed as the discovery of asymptoric freedom  \cite{GWP}. That's when the breakthrough papers of Gross, Wilczek and Politzer were submitted -- simultaneously -- to PRL. David Gross recollects \cite{DG1}:
\begin{quote}
{\em We completed the calculation in a spurt of activity. At one point a sign error in one term convinced us that [Yang-Mills] theory was, as expected, non-asymptotically free. As I sat down to put it together and to write up our results, I caught the error. At almost the same time Politzer finished his calculation and we compared, through Sidney, our results. The agreement was satisfying.}
\end{quote}

It took a few extra months for QCD to take off  as {\em the} theory of strong interactions. The events of the summer of 1973 that led to the birth of QCD are described by 
H.~Leutwyler in Section 1.1 of this Volume. To my mind, the final acceptance came with the November Revolution of 1974 -- the discovery of $J/\psi$ and its theoretical interpretation as ortho-charmonium.\footnote{I should also mention a highly motivating argument due to S. Weinberg who proved \cite{SWe} that
(in the absence of the U(1) gluon anomaly) $m_{\eta^\prime} \leq \sqrt{3} m_\pi$. This argument seemingly was discussed during  ICHEP 74 in July 1974.} In the fall of 1973 we submitted a paper \cite{BNS} explaining why the Landau theorem in four dimensions fails only in Yang-Mills theory.

QCD and its relatives  are special because QCD is the theory of {\em nature}. QCD is strongly coupled in the infrared domain  where it is impossible to treat it  quasiclassically -- perturbation theory fails even qualitatively. It does not capture drastic rearrangement of the vacuum structure related to confinement. The Lagrangian is defined  at short distances in terms of gluons and quarks, while at large distances of 
the order of $\gsim \Lambda^{-1}_{\rm QCD} $  (where $\Lambda_{\rm QCD} $  is the dynamical scale of QCD,
which I will refer to as $\Lambda$ below) we deal with hadrons, e.g. pions, $\rho$ mesons,  protons, etc. Certainly, the latter are connected with quarks and gluons in a divine way, but this connection is highly nonlinear and  non-local; even now, 50 years later, the full analytic solution of QCD is absent. 

Non-perturbative methods were desperately needed.

\section{Inception of non-perturbative methods}
\label{sectwo}

Four years before QCD Ken Wilson published a breakthrough paper \cite{KW} on the operator product expansion (OPE) whose pivotal role in the subsequent development of HEP theory 
was not fully appreciated until much later. What is now usually referred to as Wilsonian renormalization group (RG), or Wilsonian RG flow, grew from this paper. Wilsonian paradigm of separation of scales in quantum theory  was especially suitable for asymptotically free theories.
Wilson's formulation makes no reference to perturbation theory, it has a general nature and is applicable in non-perturbative regime too. The focus of Wilson's work was on statistical physics, where the program is also known as the block-spin approach. Starting from microscopic degrees of freedom at the shortest distances $a$, one ``roughens" them, step by step, by constructing a sequence of effective (composite) degrees of freedom at distances $2a$, $4a$, $8a$, and so on. At each given step $i$ one constructs an effective Hamiltonian, which fully accounts for dynamics at distances shorter than $a_i$ in the coefficient functions.

QCD required a number of specifications and adjustments. Indeed, the UV fixed point in QCD is at $\alpha_s =0$; hence, the approach to this fixed point at short distances is very slow, logarithmic rather than power-like characteristic  for the $\alpha_s \neq 0$ fixed points.  In fact, it is not the critical regime at the UV fixed point {\em per se} we are interested in but rather the regime of approach to this critical point. Moreover, it was not realized that (in addition to the dynamical scale $\Lambda$) the heavy quarks provide an extra scale --  the heavy quark mass  $m_Q$ -- which must be included in OPE where necessary. 

Surprisingly, in high-energy physics of the 1970s the framework of OPE was narrowed down to a very limited setting. On the theoretical side, it was discussed almost exclusively in perturbation theory. On the practical side, its applications were mostly narrowed down to deep inelastic scattering, where it was customary to work in the {\em leading}-twist approximation.
\label{hqm}

The fact that the UV fixed point is at zero makes OPE both more simple and more complicated than in the general case. On one hand, the anomalous dimensions of all composite local operators which might be relevant in the given problem scale only logarithmically. On the other hand,  slow  (logarithmic)
 fall off  of ``tails"  instead of desired power-like -- makes  analytic separation of scales technically difficult.

I believe that we -- Arkady Vainshtein, Valentin Zakharov and myself -- were the first to start constructing a QCD version of OPE. The first step in this direction was undertaken  in 1974 in the problem of strangeness-changing weak decays \cite{peng} (currently known as the penguin mechanism in flavor-changing decays). A mystery of 
$\Delta I=\frac 1 2$ enhancement in $K$ decays had been known for years (for a review see \cite{VA}). A suggestion of how one could apply OPE to solve this puzzle was already present in  Wilson's paper \cite{KW}. Wilson naturally lacked particular details of QCD. The first attempt to implement Wilson's idea in QCD was made in
\cite{MGBL}. Although these papers were inspirational, they missed the issue of a ``new" OPE  needed for QCD realities. Seemingly, we were the first to address this challenge, more exactly two of its features: mixed quark-gluon operators (in \cite{peng} we introduced  $$O_{\rm peng} =\bar{s}_L\gamma^\mu \left( {\mathcal D}_\nu  G^{\mu\nu} \right) d_L$$ which is purely $\Delta I=\frac 1 2$) and coefficients
logarithmically depending on the charmed (i.e. heavy at that time) quark mass. Currently, $c,b,t$  quark masses appear in the penguin operators (illustrated in Fig. 1), the latter two being genuinely heavy. Through equations of motion the operator $O_{\rm peng}$ reduces to a four-quark operator but  its chiral structure  is different from conventional, namely,  it contains both left-handed and right-handed  quark fields since ${\mathcal D}_\nu  G^{\mu\nu} \sim \sum_q \bar{q} \gamma^\mu q$. Combined with another revolutionary finding of QCD, the extraordinary smallness of the $u$ and $d$ quark masses, $m_{u,d}\sim 5\,$MeV (see Sec.1.1.15 of this Volume),  the mixed chiral structure of the emerging four-fermion operator provides  the desired  enhancement of the $\Delta I=\frac 1 2$  amplitude. It took us over two years to fight  a succession of referees for publication of the second paper in Ref. \cite{peng}. One after another, they would repeat that mixed-chirality four-fermion operator in the considered theory was complete nonsense.
Currently, the penguin mechanism  in flavor changing
weak transitions is  a basic theoretic element for a large variety of such decays. As Vainshtein put it \cite{VA}, ``Penguins spread out but have not  yet landed."
\begin{figure}
\centerline{\includegraphics[width=3cm]{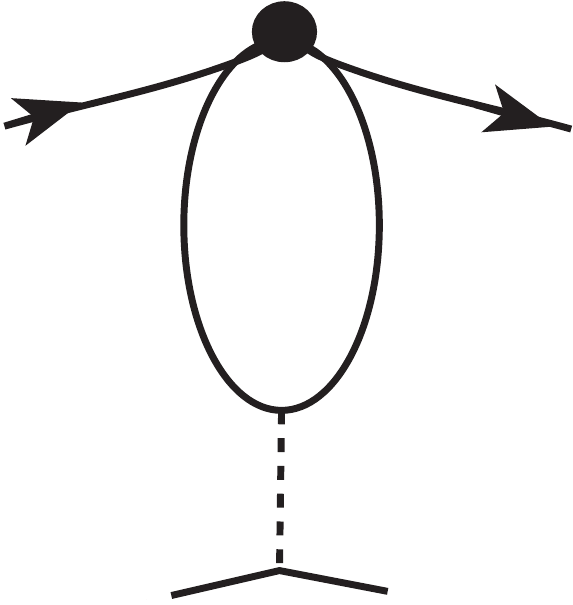}} \caption{The penguin mechanism in flavor-changing decays. Any of three heavy quarks $c,b$ or $t$ can appear in the loop.\vspace{1mm} } 
\label{peng}
\end{figure}

Systematic studies of Wilsonian OPE in QCD can be traced back to  the summer of 1977 -- that is when the the gluon condensate $O_G$ (see Table 1) was first introduced \cite{SVZ1}.
Vacuum expectation values of other gluon and quark operators were introduced in Ref. \cite{SVZ2}, which allowed one to analyze a large number of vacuum two- and three-point functions, with quite nontrivial results for masses, coupling constants, magnetic moments and other static characteristics of practically all
low-lying hadronic states of mesons and baryons. A consistent Wilsonian approach requires an auxiliary normalization point $\mu$ which plays the role of a regulating parameter separating hard contributions included in the coefficient functions and soft contributions residing in local operators occurring in the expansion. The degree of locality is regulated by the same parameter.

After  setting the foundation of OPE in QCD \cite{SVZ2} we were repeatedly returning to elaboration of various issues, in particluar, in the following works: \cite{NSVZ},\cite{Shifm2}, and \cite{Shifm3}.

\vspace{4mm}

\section{SVZ sum rules. Concepts}

The 1998 review \cite{Shifm2} summarizes for the reader foundations of the Shifman-Vainshtein-Zakharov (SVZ) sum rules in a pedagogical manner. At short distances QCD is the the theory of quarks and gluons. Yang-Mills theory of gluons confines. This means that if you have a heavy probe quark and an antiquark at a large separation a flux tube with a constant tension develops between them preventing  their ``individual" existence. In the absence of  the probe quarks the flux tube can form closed contours interpreted as glueballs. This phenomenon is also referred to in the literature as the area law or the dual Mei{\ss}ner effect. Until  1994 the above picture was the statement of faith. In 1994 Seiberg and Witten found an analytic proof \cite{SW} of 
the dual Mei{\ss}ner effect in ${\mathcal N}=2$ super-Yang-Mills.\footnote{More exactly, confinement through the flux tube formation was proven in the low-energy limit of this theory upon adding a small deformation term breaking ${\mathcal N}=2$ down to ${\mathcal N}=1$. }  The Seiberg-Witten solution does {\em not} apply to QCD, rather to its distant relative. The real world QCD, with quarks, in fact has no area law (the genuine confinement is absent) since the flux tubes break through the quark-antiquark pair creation. Moreover, light quarks are condensed, 
leading to a spontaneous breaking of chiral symmetry, a phenomenon shaping the properties of the low-lying hadronic states, both mesonic and baryonic.
The need to analytically understand these properties from first principles led us to the development of the SVZ method.

The quarks comprising the low-lying hadronic states, e.g. classical mesons or baryons, are not that far from each other, on average. The distance between them is of order
of $\Lambda^{-1}$. Under the circumstances, the string-like chromoelectric flux tubes, connecting well-separated co\-lor charges, do not  develop and details of their structure are not relevant. Furthermore, the valence quark pair injected in the vacuum -- or three quarks in the case of baryons -- perturb it only slightly. Then we do not need the full machinery of the QCD strings\,\footnote{Still unknown.} to approximately describe the properties of the low-lying states. Their basic parameters depend on how the valence quarks of which they are built interact with typical vacuum field fluctuations.

We endowed the QCD vacuum with various condensates -- approximately a half-dozen of them  -- in the hope that this set would be sufficient to describe a huge variety of the  low-lying hadrons, mesons and baryons. The original set included\,\footnote{A meticulous writer would have used the notation $\left\langle G_{\mu\nu}^2 \right\rangle$, etc.  but I will omit bra and ket symbols where there is no menace of confusion.  }
 the gluon condensate $G_{\mu\nu}^2$, the quark condensate $\bar{ q} q$, the mixed condensate $\bar q\sigma^{\mu\nu}G_{\mu\nu}q$, various four-quark condensates
 $\bar{q}\Gamma q \bar{q}\Gamma q $, 
 and a few others (see Table 1). Later this set had to be expanded to address such problems as, say, the magnetic moments of baryons.

\begin{figure}
\centerline{\includegraphics[width=8.5cm]{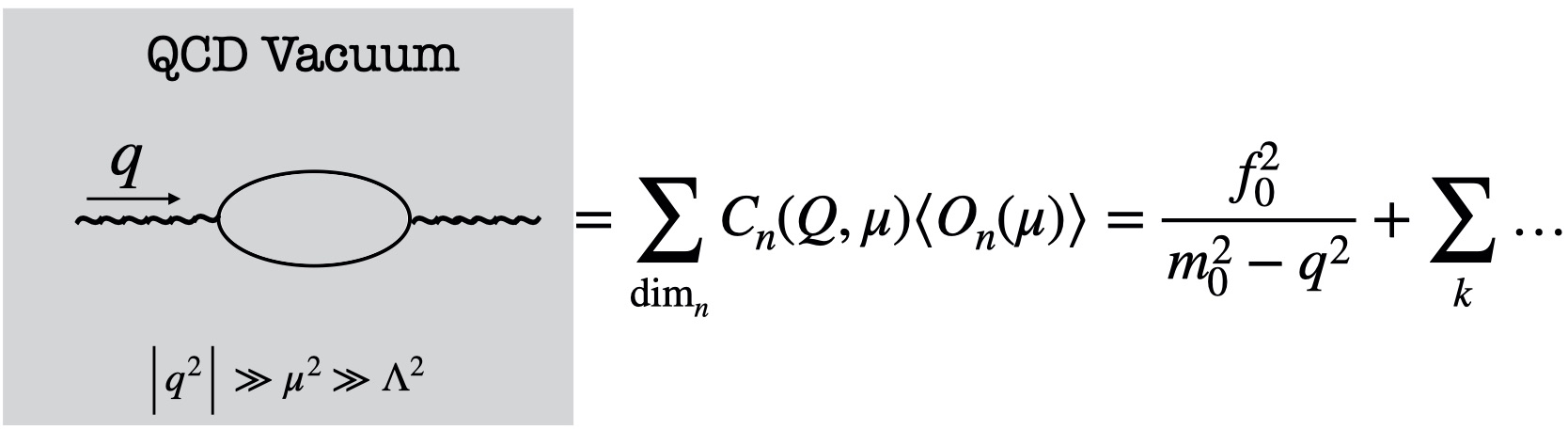}} \caption{A two-point correlation function in the QCD vacuum. The left  side is the OPE sum with a finite number of the lowest-dimension operators ordered according to their normal dimensions. The right side is the sum over  mesons with the appropriate quantum numbers. The ground state in the given channel is singled out. The excited states are accounted for in the quasiclassical approximation. We define a positive variable $Q^2=-q^2$ and a sliding 
$\mu^2$ parameter used as a separation parameter in OPE. For better convergence a Borel transformation is applied as explained below.} 
\label{sczsr}
\end{figure}

\begin{table*}[t]
\caption{The lowest-dimension operators in OPE. $\Gamma$ is a generic notation for combinations of the Dirac $\gamma$ matrices.}
\label{tab:1}      
\begin{tabular}{llllll}
\hline\noalign{\smallskip}
Normal dim  & 3 & 4 & 5 & 6 & 6\\
\noalign{\smallskip}\hline{\smallskip}
Operator \rule{0mm}{4mm}& $O_q =\bar{q}q$ & $O_G= G_{\mu\nu}^2$ & $O_{qG}=\bar{q}\sigma^{\mu\nu}G_{\mu\nu}q$ & $O_{4q} = (\bar{q} \Gamma q)^2   $ &$ O_{3G} = GGG $  
\end{tabular}
\vspace{5mm}
\end{table*}

Our task was to determine the regularities and parameters of the classical mesons and baryons from a limited set of the vacuum condensates.
Figure \ref{sczsr}
graphically demonstrates  the SVZ concept. On the theoretical side,  an appropriate $n$-point function is calculated as an OPE expansion truncated at a certain order. 
In most problems only condensates up to dimension 6 (Table 1) are retained. In the ``experimental" part the lowest-lying meson (or baryon) is singled out, while all higher states are represented in the quasiclassical approximation. Acting in this way, one can determine the parameters $f_0$ and $m_0$ defined in Fig. \ref{sczsr} and their analogs in other problems. Of course, without invoking the entire infinite set of condensates one can only expect to obtain the hadronic parameters in an admittedly approximate manner.
 
 \section{Borelization}
 
 \begin{figure}
\centerline{\includegraphics[width=5.5cm]{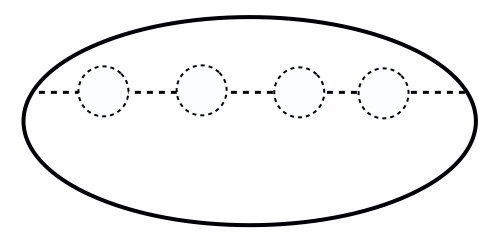}} \caption{Graph showing four loops renormalizing a gluon line (represented by the dotted line). A renormalon is the sum 
over $n$ of such diagrams with $n$ loops.
\vspace{-1mm} } 
\label{peng}
\end{figure}

Analyzing the sum rules displayed in Fig. 2 we realized that their predictive power was limited -- summation on both sides of the equation does not converge fast enough. On the right-hand side the contribution of high excitations is too large -- the lowest lying states are ``screened" -- because the weight factors fall off rather slowly. Likewise, to achieve reasonable accuracy on the left-hand side one would need to add operators other than those collected in Table 1 of which we knew next to nothing. The Borel transform came to the rescue.

The Borel transformation is a device well-known in mathematics. If one has a function $f(x)$ expandable in the Taylor series, $f(x) =x\sum_n a_nx^n$  with the coefficients $a_n$ which do not fall off sufficiently fast, one can instead introduce its Borel transform
\beq
{\mathcal B}f = x\sum_n\frac{a_n}{n!} x^n
\eeq
and then, if needed, reconstruct $f$.\footnote{The Borel transform is closely related to the Laplace transform.}
If we apply this procedure to the sum rule in Fig. 2
we obtain for a given hadronic state $i$
\vspace{1mm}

\beqn
{\mathcal B}\frac{f_i^2}{m_i^2+Q^2} &=& {\mathcal B}\,\frac{f_i^2}{Q^2}\,\sum_n(-1)^n\left[ \frac{m_i^2}{Q^2} \right]^n\nonumber\\[1mm]
&\to&
\frac{f_i^2}{Q^2}\,\sum_n\frac{(-1)^n}{n!}\left[ \frac{m_i^2}{Q^2} \right]^n\nonumber\\[1mm]
&=& \frac{f_i^2}{Q^2}\,
 \exp\left(-\frac{m_i^2}{Q^2}\right)
\nonumber\\[1mm]
&\equiv& \frac{f_i^2}{M^2}\left(-\frac{m_i^2}{M^2}\right)
\label{borppp}
\eeqn
where in the final step (for historical reasons)  I replaced $Q^2$ by a Borel parameter $M^2$. If $M^2$ can be chosen sufficiently small, higher excitations are {\em exponentially} suppressed. 

Simultaneously, we improve the convergence of  OPE on the left-hand side by applying the same operator. If the operator $O_n$ has dimension $2d_n$
then the Borel transformation of the left hand side yields
\beq
  \sum_n \frac{1}{(Q^2)^{d_n}}\langle O_n\rangle \,\to \,   \sum_n\frac{1}{(d_n-1) !}  \frac{1}{(M^2)^{d_n}} \langle O_n\rangle\,,
  \label{41}
\eeq

\vspace{1mm}

where I have again 
replaced $Q^2$ by the Borel parameter $M^2$. Since the expansion (\ref{41}) goes in inverse powers of $M^2$, it is necessary to keep $M^2$  large enough. The two requirements on $M^2$  seem contradictory. However, for all ``typical" resonances, such as, say, $\rho$ mesons, they can be met simultaneously \cite{SVZ2} in a certain ``window." The only exception is the $J^P = 0^\pm$ channel. There are special reasons why $0^\pm$ mesons are exceptional.
  
   \section{Practical version of OPE}
   
   At the early stages of the SVZ program the QCD practitioners often did not fully understood the concept of the scale separation in the Wilsonian OPE. It was generally believed that
 the coefficients are fully determined by perturbation theory while non-perturbative effects appear only in the OPE operators.\footnote{Unfortunately, this misconception lasted through the 1980s and was visible in the literature even in the 1990s and later.} This belief led to inconsistencies which revealed themselves e.g. in the issue of renormalons (see below). A set of graphs represented by renormalons is constructed from a single gluon exchange by inserting any number of loops in the gluon line like beads in a neckless (e.g Ref. \cite{Shifm3} and references therein). Being treated formally this contribution, shown in Fig. 3, diverges factorially at high orders.
 I vividly remember that after the first seminar on SVZ  in 1978 Eugene Bogomol'nyi asked me each time we met: ``Look, how can you speak of power corrections in the $n$-point functions at large $Q^2$ if even the perturbative expansion (i.e. the expansion in 
 $1/\log(Q^2/\Lambda^2)$) is not well defined? Isn't it inconsistent?" I must admit that at that time my answer to Eugene  was somewhat evasive.

 The basic principle of Wilson's OPE -- the scale separation  principle -- is ``soft versus hard" rather than ``perturbative versus non-perturbative." Being defined in this way the condensates are explicitly $\mu$  dependent. All physical quantities are certainly $\mu$  independent; the normalization point dependence of the condensates is compensated by that of the coefficient functions, see Fig. 2. 
  
  The problem of renormalons disappears simultaneously with the introduction of the normalization point $\mu$. With $\mu\gg\Lambda$, there is no factorial divergence 
  in high orders of perturbations theory. Renormalons conspire with the gluon condensates to produce, taken together, a well-defined OPE. The modern construction goes under the name of the ``renormalon conspiracy"; it is explained in detail in my review \cite{Shifm3}. I hasten to add, though,  that the renormalons acquire a life of their own in those cases in which OPE does not exist. Qualitatively, they can shed light on scaling dimensions of non-perturbative effects. The most clear-cut example of this type is the so-called  ``pole mass of the heavy quarks" \cite{Shifm7p} and its relation to a theoretically well-defined mass parameter \cite{Shifm6}.
  
  In some two-dimensional solvable models exact OPE can be constructed which explicitly demonstrates the $\mu$ dependence of both the coefficient
  functions and the condensates in the Wilsonian paradigm  and its cancellation in the physical quantities (for a recent study see e.g. \cite{sch}). Needless to say, if QCD was exactly solved we would have no need in the SVZ sum rules. 
  
  We had to settle for a reasonable compromise, known as {\em the practical version of OPE}. In the practical version we calculate the coefficient functions perturbatively keeping a limited number of loop corrections. The condensate series is truncated too. The condensates are not calculated from first principles; rather a limited set is determined from independent data.
  
The practical version is useful in applications only provided $\mu^2$ can be made small enough to ensure that the ``perturbative" contributions to the condensates are much smaller than their genuine (mostly non-perturba\-tive) values. At the same time, $\alpha_s( \mu^2 )/\pi$  must be small enough for the expansion in the coefficients  to make sense. The existence of such\, ``$\mu^2$ window" is not granted {\em a priori} and is a very fortunate feature of QCD. We did observe this feature empirically 
in almost  all low-lying hadrons \cite{vol}.~\footnote{Theoretical understanding of the roots of this phenomenon remains unclear. Seemingly, it  has no known analogues in two-dimensional models. }  At the same time, we identified certain exceptional channels revealing unforeseen nuances in hadronic physics \cite{alike}. 

\vspace{-4mm}

   \section{Implementation of the idea and results}
 
After the strategic idea of quark and gluon interaction with the vacuum medium became clear we delved into the uncharted waters of microscopic hadronic physics. Remember, in 1977 nobody could imagine that  basic hadronic parameters for at least some hadrons could be analytically calculated, at least approximately. As a show-case example we chose the most typical mesons,  $\rho$ and $\phi$, to calculate their couplings to the electromagnetic current and masses. The agreement of our results with experiment was better than we could {\em a priori} expect. At first we were discouraged by a wrong sign of the gluon condensate term in the theoretical part of the appropriate SVZ sum rule. We suddenly understood that this sign could be compensated by the four-quark condensate -- a real breakthrough. In November of 1977 we published a short letter \cite{SVZ1}
which still missed a number of elements (e.g. Borelization) developed and incorporated later, one by one. We worked at a feverish pace for the entire academic year, accumulating 
a large number of results for the hadronic parameters. All low-lying meson resonances built from the $u,d,s$ quarks and gluons were studied and their static properties 
determined from SVZ: masses, coupling constants, charge radii, $\rho$-$\omega$ mixing,  and so on, with unprecedented success. In summer of 1978, inspired by our progress, we prepared a number of preprints (I think, eight of them simultaneously\,\footnote{In the journal publication they were combined in three articles occupying the whole issue of Nucl. Phys. B147, ${\mathcal N}^{\underline{o}} 5$, see \cite{SVZ2}.}) and submitted to ICHEP-78 in Tokyo. Seemingly none of us were allowed to travel to Tokyo to present our results. 

I cannot help mentioning an incident that occurred in the spring of 1978 when
we were mostly done with this work. The episode may have been funny were it not so nerve-wracking. When we decided that the calculational  stage of the work was over, I collected all my drafts (hundreds of sheets of paper with derivations and math expressions), I organized them in proper order, selected all expressions we might have needed for the final draft of the paper and the future work, meticulously rewrote them in a voluminous notebook (remember, we had no access to photocopying machines), destroyed the original drafts, put the the notebook in my briefcase and went home. It was about midnight, and I was so exhausted that I fell asleep while on the metro train. A loud voice announcing my stop awoke me, and I jumped out of the train, leaving the briefcase were it was, on the seat. By the time I realized what have happened the train was gone, and gone with it forever my calculations ... I have never recovered my briefcase with the precious notebook... After a few agonizing days it became clear that the necessary formulas and expressions had to be recovered anew. Fortunately, Vainshtein and Zakharov  had kept many of their own derivations. Vainshtein never throws away anything as a matter of principle. Therefore, the problem was to dig out ``informative" sheets of paper from the ``noise" (this was hindered by the fact that Vainshtein was in Novosibirsk while we were in Moscow). Part of my drafts survived in the drawers of a huge desk that I had inherited from V. Sudakov. Better still, many crucial calculations were discussed so many times by us, over and over again, that I remembered them by heart. Nevertheless, I think it took a couple of uneasy weeks to reconstruct in full the contents of the lost notebook.

The SVZ method was further developed by many followers (e.g. the so-called light-come sum rules for form-factors), see \cite{Braun} and \cite{Khodj}. A broad picture of the hadronic world was obtained by the 1980s and later \cite{Colangelo}. Today the pioneering SVZ paper is cited 6000+ times. Until 1990s, when lattice QCD based on numeric calculations, started approaching its maturity,
the SVZ method was the main tool for analyzing static hadronic properties. 

\vspace{-4mm}

   \section{Reliability  and predictive power}
   \label{secseven}
   
   The SVZ method is admittedly approximate. Yet, it is not a model in the sense that it cannot be arbitrarily bent to accommodate ``wrong" data. It is instructive to narrate here the story of an alleged discovery of an alleged  ``paracharmonium" referred to as $X(2.83)$  in January of 1977 \cite{X}. It was widely believed then that $X(2.83)$ was the $0^-$ ground sate of $\bar c c$ quarks, $\eta_c$. If this was the case the mass difference between $J/\psi$ and $\eta_c$ would be close to 270 MeV. Shortly after, the interpretation 
   of $X(2.83)$  as $\eta_c$ was categorically ruled out by the SVZ analysis \cite{volo} which {\em predicted}  that the above mass difference must be 100$\pm$ 30 MeV.
  Two years later, a new experiment \cite{par} negated the existence of the  $X(2.83)$ state.  In the very same experiment the genuine paracharmonium was observed at $2.98\pm 0.01$ GeV, in perfect agreement with \cite{volo}. For us this was a triumph and a lesson -- if one believes in a theory one should stand for it! \label{etasubc}
  
      \section{OPE-based construction of heavy quark mass expansion}
      
      In the 1980s and early 1990s OPE was generalized to cover theoretical studies of  mixed heavy-light hadrons, i.e. those built from  light, $q$, and heavy, $Q$, flavors.  In the 1990s  those who used $1/m_Q$ expansion in theoretical analysis of $Q\bar q$ and $Qq q$ systems numbered in the hundreds. A large range of practical physics problem related to $Q\bar q$ and $Qqq$ systems were solved. Lattice analyses of such systems 
even now remain hindered, and in many instances the $1/m_Q$ expansion remains the only reliable theoretical method.

\begin{figure}
\centerline{\includegraphics[width=8cm]{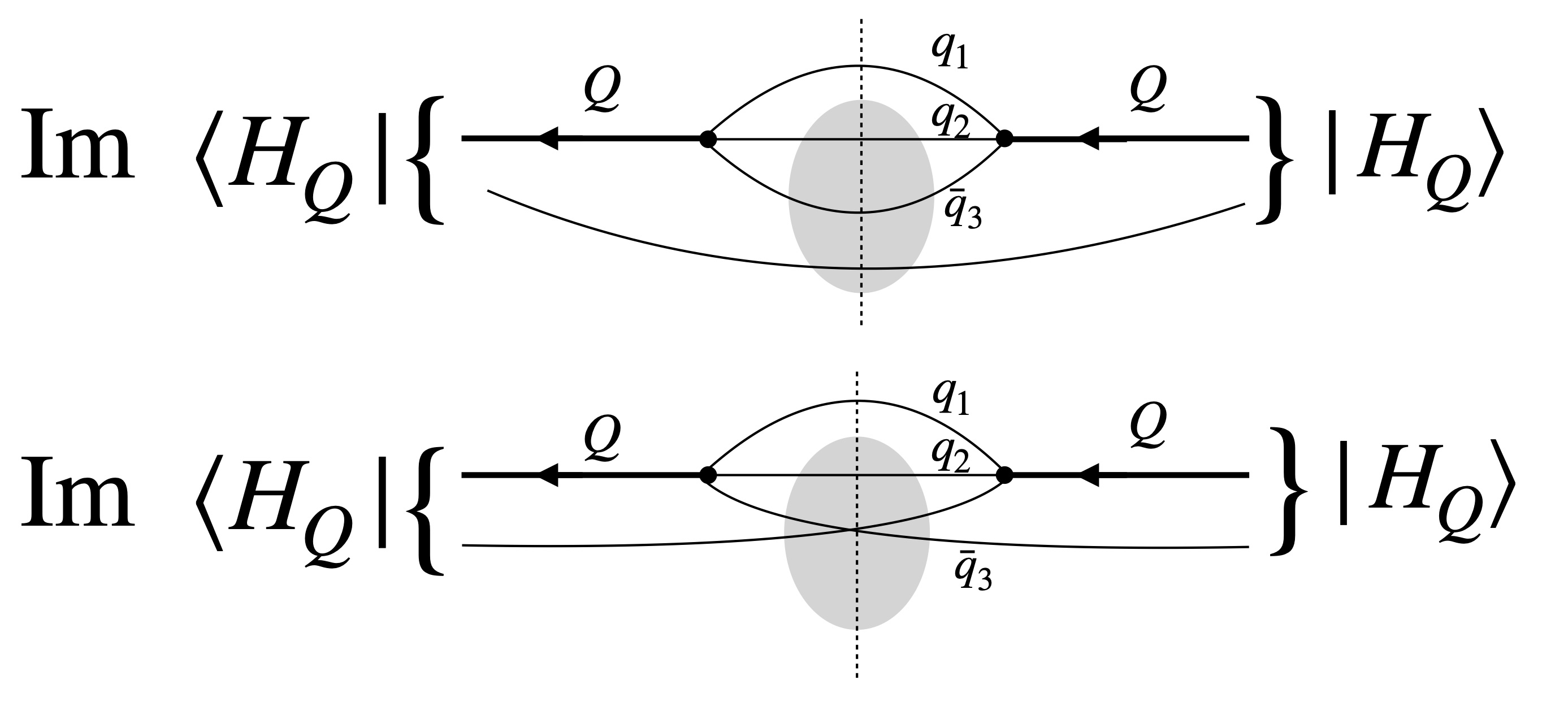}} \caption{ 
$1/m_Q$ expansion for a $H_Q$ weak inclusive decay rate (see Eq. (\ref{hqidr})). Depicted are two operators, the leading 
$\bar{Q} Q$ and a subleading $(\bar{Q} q_3) (\bar{q}_3 Q) $.  Both are sandwiched between the heavy hadron states $\langle H_Q |$ and $|H_Q\rangle$ and the decay rate is determined by the imaginary part.
The grey area depicts the soft quark-gluon cloud. Adapted from Ref. \cite{Sh1}. 
} 
\vspace{5mm}
\label{hqp}
\end{figure}

As I have mentioned in the second paragraph of Sec. \ref{sectwo}, heavy quarks in QCD introduce  an extra scale, $m_Q$. To qualify as a heavy quark $Q$ the corresponding mass term $m_Q$ must be much larger than $\Lambda_{\rm QCD}$. The charmed quark $c$ can be considered as heavy only with some reservations while $b$ and $t$ are {\em bona fide} heavy quarks.  The hadrons composed from one heavy quark $Q$, a light antiquark $\bar q$, or a ``diquark" $qq$, plus a gluon cloud (which also contains light quark-antiquark pairs) -- let us call them $H_Q$ -- can be treated in the framework of OPE. The role of the cloud is, of course, to keep all the above objects together, in a colorless bound state. The light component of $H_Q$, its light cloud, has a complicated structure; the soft modes of the light fields are strongly coupled and strongly fluctuate. Basically, the only fact which we know for sure is that the light cloud is indeed light; typical excitation frequencies are of order of $\Lambda$. One can try to visualize the light cloud as a soft medium.\footnote{Hard gluons do play a role too. They have to be taken into account in the coefficient functions as will be mentioned In Sec. \ref{hagl}.}  The heavy quark $Q$ is then submerged in this medium. The latter circumstance allows one to develop a formalism similar to SVZ in which the soft QCD vacuum medium is replaced by that of the light cloud. As a result, an OPE-based expansion in powers of $1/m_Q$ emerges (see Fig. \ref{hqp}). When heavy quarks are in soft medium
the heavy quark-antiquark pair creation does not occur and the field-theoretic description of the heavy quark becomes redundant. A large ``mechanical" part in the $x$ dependence of $Q(x) $ can be {\em a priori} isolated, $Q(x) = \exp({-im_Qt} )\tilde{Q}({x})$. The reduced bispinor field $ \tilde{Q}(x)$ describes a residual heavy quark motion inside the soft cloud; the heavy quark mass appears only in the form of powers of $1/m_Q$ (first noted in 1982).

Comprehensive reviews on the OPE-based heavy quark theory exist \cite{Shifm6},\cite{ur},\cite{Shifm7},\cite{hqr}. There the reader will find exhaustive lists of references to original publications.  Therefore, in my presentation below I will be brief, with a focus on a historical aspect, as I remember it, and limit myself to a few selected references.

In the early 1980s abundant data on the meson and baryon $H_Q$ states started to appear. Theoretical understanding of the total decay rates beyond the free-quark calculations became a major goal. This challenge paved the way to the beginning of the $1/m_Q$ expansion in $H_Q$ hadron physics in the mid 1980s.
The decay rate 
into an 
inclusive final state $f$ can be written in terms of the imaginary part of a  forward
scattering operator (the so-called transition operator)  evaluated to
second order in the  weak interactions \cite{Sh1},
\begin{equation} 
{\rm Im} \hat T(Q\!\rightarrow\! f\!\rightarrow\! Q)\! = \!
 {\rm Im} \!\int d^4x\ i\,T \left({\cal L}_W(x){\cal 
L}_W^{\dagger}(0)\right)\ 
\label{OPTICAL} 
\end{equation} 
where $T$ denotes the time ordered product and 
${\cal L}_W$ is the relevant weak Lagrangian at the normalization 
point  $\mu \sim m_Q$. The factor $\exp({-im_Qt} )$ mentioned above is implicit in Eq. (\ref{OPTICAL}).
Descending to $\mu \ll m_Q$ one arrives at the OPE expansion 

\vspace{1mm}

\beqn
&&\Gamma (H_Q\rightarrow f) = G_F^2 |V_{\rm CKM}|^2m_Q^5   
\sum _i  \tilde c_i^{(f)}(\mu ) 
\frac{\matel{H_Q}{O_i}{H_Q}_{\mu }}{2M_{H_Q}}  \nonumber\\[1mm]
&&  
\propto  \left[ c_3^{(f)}(\mu )\frac{\matel{H_Q}{\bar 
QQ}{H_Q}_{(\mu)}}{2M_{H_Q}} \right.\nonumber\\[1mm]
&&\left. + c_5^{(f)}(\mu ) m_Q^{-2}
\frac{
\matel{H_Q}{\bar Q\frac{i}{2}\sigma G Q}{H_Q}_{(\mu )}}
{2M_{H_Q}\;\;} \right. \nonumber\\[1mm]
&&\left. +\sum _i c_{6,i}^{(f)}(\mu )m_Q^{-3}\frac{\matel{H_Q}
{(\bar Q\Gamma _iq)(\bar q\Gamma _iQ)}{H_Q}_{(\mu )}}
{2M_{H_Q}\;} \right.\nonumber\\[1mm]
&&\left.+ {\cal O}(1/m_Q^4)+ ...\right]   ,
\label{hqidr}
\eeqn
where $\Gamma_i$ represent various combinations of the Dirac $\gamma$ matrices, see also table 1.
In SVZ we dealt with the vacuum expectation values of relevant operators while in the heavy quark physics the relevant operators are sandwiched between $H_Q$ states.

\vspace{-2mm}

\section{Applications}

The expansion (\ref{hqidr}) allowed us to obtain \cite{Sh1} the first {\em quantitative} predictions for the hierarchies of the lifetimes of $Q\bar q$ mesons and $Q q q$ baryons
($Q$ was either $c$ or $b$ quark)  in the mid-1980s  -- another spectacular success of the OPE-based methods. The dramatic story of $\eta_c$ narrated in Sec. \ref{secseven}
repeated itself. With the advancement of experiment in the late 1990s, a drastic disagreement was allegedly detected in the ratio $\tau(\Lambda_b)/\tau(B_d)_{\rm exp} =0.77\pm 0.05$
compared to the  theoretical prediction $$\tau(\Lambda_b)/\tau(B_d)_{\rm theor} = 0.9\pm 0.03$$ (e.g. \cite{Shifm6}). In the 2010s the $\Lambda_b$ lifetime was remeasured shifting the above experimental ratio up to $0.93\pm 0.05$. Hurrah!

In the mid-1980s, at the time of the initial theoretical studies of the $H_c$ and $H_b$ lifetime hierarchies \cite{Sh1}, 
next to nothing was known about heavy baryons. 
Since then enormous efforts were invested in improving theoretical accuracy both in {\em  mesons and baryons} in particular by including higher-dimension operators in the  inverse heavy quark mass expansion (IHQME)  and higher-order $\alpha_s$ terms  in the OPE coefficients.
The status of the IHQME for $H_Q$ lifetimes as of 2014 was presented in the review \cite{Lenz:2014jha}. The advances reported there and in more recent years cover more precise determination of the matrix elements of four-quark operators via HQET sum rules \cite{Kirk:2017juj}, calculations of the higher $\alpha_s$ corrections, in particular, 
$\alpha_s^3$ corrections to the semileptonic $b$ quark decay \cite{Fael:2020tow},  the first determination of the Darwin coefficient for non-leptonic decays \cite{Lenz:2020}, etc.  Comparison with the current set of data on $\tau({H_c})$ can be found in \cite{Rusov}. In this context I should also mention an impressive publication 
\cite{Gratrex:2022xpm} (see also references therein) which, in addition to a comprehensive review of the OPE-based analysis of the $H_c$ lifetimes, acquaints the reader with a dramatic story of the 
singly charmed  {\em baryon} hierarchy. Indeed, according to PDG-2018 the lifetime of $\Omega_c^0$ is $69\pm 12\,$fs while PDG-2020 yields $\tau (\Omega_c^0) = 268\pm 24\pm\! 10\!$ fs! The latter measurement  was reported by LHCb \cite{LHC}. Moreover, the newest Belle II result \cite{belle} which has been published in August 2022,
$\tau (\Omega_c^0)=243 \pm 48 \pm 11\!$ fs, triumphantly confirms the LHCb data.
The jump in the $\Omega_c^0$ lifetime by a factor of 3 to 4 compared to the older measurements  could happen only because (presumably) statistical and/or systematic errors in the older measurements  were grossly underestimated.
With these fresh data the observed hierarchy of lifetimes changes -- $\Omega_c^0$ from the first place (the shortest living $H_c$ baryon)  moves to the third.\footnote{ It is curious to note that 30 years ago Blok and I argued 
\cite{blok} (Secs. 4.2 and 6) that $\Omega_c^0$ could be the longest living singly charmed  baryon due to its $ss$ spin-1 diquark structure.}  The question arises whether the OPE-based theory can explain the current experimental hierarchy $\tau (\Xi^0 ) < \tau(\Lambda_c^+ )
< \tau (\Omega_c^0 ) <  \tau (\Xi^+)$. In \cite{Gratrex:2022xpm} it is argued that the answer is ``yes, it is possible" (see Fig. 5 in \cite{Gratrex:2022xpm})  provided one takes into account $1/m_c^4$ contributions due to four-quark operators and $\alpha_s$ corrections in the appropriate coefficient functions.\footnote{The four-quark operators introduced in \cite{Sh1} responsible for the Pauli interference yield corrections $O(1/m_c^3$), see Eq. (\ref{hqidr}). The authors of \cite{Gratrex:2022xpm} go beyond this set.} 

I should emphasize that the theoretical accuracy in the $H_c$ 
family is limited by the fact that the expansion parameter $\Lambda_{\rm QCD}/m_c$ is not small enough.  Even including sub-leading contributions will hardly provide us with high-precision theoretical predictions. For $H_c$ states IHQME at best provides us with  a semi-quantita\-tive guide. On the other hand, in the theory of $H_b$ decays one expects much better accuracy.

\vspace{-5mm}

       \section{Around 1990s and beyond}
       
   \subsection{Heavy quark symmetry in the limit \boldmath{$m_Q\to \infty$}}

The light-cloud interpretation as in Fig. \ref{hqp}  immediately implies  that at zero recoil the (appropriately normalized) $B\to  D$ formfactors reduce to unity.
  This is called the ``small velocity (SV) limit theorem" \cite{NW}, \cite{VS}.   The above``unification"  is similar to the vector charge non-renormalization theorem at zero momentum transfer, say, for the $\bar u\gamma^\mu d$ current. 
  The $D$ and $B$ masses are very far from each other. One has to
subtract the
mechanical part
of the heavy quark mass in order to see that all dynamical
parameters
are insensitive to the substitution $Q_1\leftrightarrow Q_2$
in the limit $m_{Q_{1,2}}\rightarrow\infty$, with the SV limit ensuing at zero recoil. Perhaps,
this is the
reason why it was discovered so late. The next step was made by Isgur and Wise who generalized this symmetry off the zero-recoil point by virtue of the 
 Isgur-Wise function \cite{IW}. 
 
 \vspace{-7mm}
       
\subsection{HQET}
  
  Heavy quark effective theory which emerged in the 1990s \cite{ee} formalizes and automates a number of aspects of the generic $1/m_Q$ expansion. 
  In fact, it immediately follows from the construction similar to (\ref{hqidr}). Simplified rules of behavior proved to be very helpful for QCD practitioners in the subsequent development of various applications.
  In HQET the reduced field 
  $\tilde Q$ is treated quantum-mechanically, its non-relativistic nature is built in, and  the normalization point $\mu$ is $\ll m_Q$ from the very beginning.\footnote{ I personally prefer to consider the heavy quark expansions directly in
{\em full QCD} in the framework of
the Wilson OPE bypassing the intermediate stage of HQET.}
Applying the Dirac equation to eliminate  small (lower) components of the Dirac spinor in favor of the large components
it is easy to derive the expansion of ${\cal L}_{\rm heavy}^0$, up to terms $1/m_Q^2$, 
\begin{align}
  &{\cal L}_{\rm heavy}^0 = 
  \bar Q (i\not\!\!D -m_Q)Q \nonumber\\[1mm]
  &=\bar Q
  \frac{1+\gamma_0}{2}\left(1+\frac{(\bfsigma\bfpi)^2}{8m_Q^2}
  \right)\left[
  \pi_0-\frac{1}{2m_Q}(\bfpi\bfsigma)^2\;-\right. \nonumber\\[1mm]
  &\left.
    -\;\frac{1}{8 m_Q^2}\,\left(-(\vec D\vec E)+2\bfsigma\cdot\vec
    E\times\bfpi\right)\,
    \right]\left(1+\frac{(\bfsigma\bfpi)^2}{8m_Q^2}
    \right)\frac{1+\gamma_0}{2}\,Q \nonumber\\[1mm]
    &+\;{\cal O}\left(\frac{1}{m_Q^{3}}\right)\, ,
    \label{HQLm}
\end{align}
where $\bfsigma$ denote the Pauli matrices and
$$
({\bfpi}\bfsigma )^2 ={\bfpi}^2 +\bfsigma \vec B\, ,
$$
$\vec E$
and $\vec B$ denote the background chromoelectric and
chromomagnetic fields, respectively. Moreover, the operator $\pi_\mu$ is defined through
\beqn
iD_\mu Q(x) &&= e^{-im_Qv_\mu x_\mu}\left( m_Q v_\mu + i
D_\mu\right) \tilde Q (x) \nonumber\\[1mm]
&&\equiv e^{-im_Qv_\mu x_\mu}\left( m_Q v_\mu + \pi_\mu
\right) \tilde Q (x)
\eeqn
where $v_\mu$ is the heavy quark four-velocity.

The set of operators presented in (\ref{HQLm}) plays the same basic role in  $1/m_Q$ expansion as the set in Table 1 in SVZ  sum rules.

In the remainider of this section I will briefly mention some classic problems with heavy quarks which were successfully solved in the given paradigm.

\vspace{-3mm}

\subsection{CGG/BUV theorem}

Up to order $1/m_Q^2$ all inclusive decay widths of the $H_Q$ mesons are universal and coincide with the parton-model result for the $Q$ decay \cite{chay}, \cite{BUV},
\beq
\Gamma = \Gamma_0\left( 1-\frac{\mu_\pi^2}{2m_Q^2}\right), \qquad \mu_\pi^2 = \frac{1}{2M_{H_Q}}
\langle H_Q|\bar Q {\vec\pi}^2
Q|H_Q\rangle 
\label{theorem}
\eeq
where $\Gamma_0$ is the parton model result. There are no corrections $O(1/m_Q)$. This statement is known in the lite\-rature as the
CGG/BUV  theorem.

\subsection{Spectra and line shapes}

Lepton spectra in semileptonic $H_Q$  decays were derived in \cite{Bigi}. The leading corrections arising at  the $1/m_Q$ level were completely expressed in terms of the difference in the mass of $H_Q$ and $Q$. Nontrivial effects appearing at the order $1/m_Q^2$  were shown to affect mainly the endpoint region; they are different for meson and baryon decays as well as for beauty and charm decays.

The
theory of the
line shape in $H_Q$ decays, such as $B\rightarrow X_s \gamma$ where $X_s$ denotes
the inclusive hadronic state with the $s$ quark, resembles that of the M\"{o}ssbauer
effect.  It is absolutely remarkable
that for 10 years there were no attempts
to treat the spectra and line shapes  along essentially the same
lines as it had been done in deep inelastic scattering (DIS) in the 1970s.
Realization of this fact came only  in 1994;  technical  implementation of the idea was carried out in
\cite{JR}, \cite{motion-}, and \cite{motion}.

\subsection{Hard gluons}
\label{hagl}

\vspace{-3mm}

Hard-gluon contributions special for the heavy quark theory result in powers of the logarithms  $\alpha_s\log\left(m_Q/\mu\right)$. They determine the coefficients $c_i$ in Eq. (\ref{hqidr}) through the anomalous dimensions of the corresponding operators. They were discovered in \cite{vsp} and were called the {\em hybrid} logarithms. In HQET they are referred to as matching logarithms.

\vspace{-3mm}

\subsection{In conclusion}

Concluding the heavy quark portion I should add that Kolya Uraltsev (1957-2013), one of the major contributors in heavy quark theory  died in 2013 at the peak of his creative abilities (see \cite{ShifmanKolya}).

Concerning the OPE-based methods in QCD in general, I would like to make an apology to the many authors whose works have not been directly cited. The size limitations are severe. The appropriate references are given in the review papers listed in the text above.

       Just for the record, a couple of reviews which are tangentially connected to the topic of the present article are given in Refs. \cite{Shifm4} and  \cite{Shifm5}.
       
       \vspace{-3mm}
       
        \section{Recent developments unrelated to the OPE-based methods}      
 
 Quantum field theories from the same class as QCD are now experiencing dramatic changes and rapid advances in a deeper understanding of anomalies. I want to mention  two crucial papers: \cite{Gaiotto:2014kfa} and \cite{Gaiotto:2017yup}. The latter demonstrates  that at $\theta=\pi$ there is a discrete 't Hooft anomaly involving time reversal and the center symmetry. It follows that at $\theta=\pi$ the vacuum cannot be a trivial non-degenerate gapped state.
  \vspace{-3mm}

\section*{Acknowledgments}

This work is supported in part by DOE grant DE-SC0011842.

I am grateful to Franz Gross, Alexander Khodjami\-rian, Alexander Lenz and Bla\v{z}enka Meli\'c for very useful discussions and comments.

\end{document}